\documentstyle[aps,epsfig,twocolumn,citesort,psfig19,floats]{revtex}

\newcommand{\aleq}{\mbox{\ 
\raisebox{-.9ex}{$\stackrel{\textstyle<}{\sim}$}\ }}
\newcommand{\ageq}{\mbox{\
\raisebox{-.9ex}{$\stackrel{\textstyle >}{\sim}$}\ }}

\def\cinf{{c_{\epsilon , \infty}}}
\def\la{{\langle}}
\def\ra{{\rangle}}

\def\eps{{\epsilon}}

\def\rel{{Re_\lambda}}

\def\begineq{\begin{equation}}
\def\endeq{\end{equation}}
\def\be{\begin{equation}}
\def\ee{\end{equation}}

\begin{document}
\bibliographystyle{prsty}
%\psdraft

\title{
Periodically kicked turbulence
}
\author{
Detlef Lohse 
}
\address{
Department of Applied Physics  and J.\ M.\ Burgers Centre for
Fluid Dynamics, 
University of Twente, 7500 AE Enschede, 
Netherlands}

\date{\today}

\maketitle
\begin{abstract}
Periodically kicked turbulence is theoretically analyzed within a
mean field theory.
For large enough kicking strength $A$ and kicking frequency $f$
the Reynolds number 
grows exponentially and then runs into some saturation. The saturation
level 
$Re^{sat}$ 
can be calculated
analytically;
different regimes can be observed. 
For large enough $Re$ we find  $Re^{sat} \propto 
Af$, but intermittency can modify this scaling law. 
We suggest an experimental realization of periodically kicked turbulence to
study the different regimes we theoretically predict and thus to better
understand the effect of forcing on fully developed turbulence. 
\end{abstract}

%----------------------------------------------------------------------

\vspace{1cm}

Periodically driven flow is ubiquitous. Faraday's experiment
\cite{far31} is an idealized version thereof, more relevant examples are the
earth's atmosphere, driven by periodical heating of the sun, or 
the blood flow in veins, driven by the beating heart. Another example
is the gas flow inside a sonoluminescing bubble
which is periodically kicked by the collapsing bubble 
wall
\cite{cru94}.
Another example of periodically kicked turbulence is 
the numerical realization of homogeneous shear flow 
\cite{pum96} where periodical remeshing is necessary.

In this paper we set up a mean field theory for periodically kicked flow, based
on the mean field theory for decaying turbulence 
\cite{loh94a} which was able to
describe the experimentally measured energy decay in turbulent liquid
helium flow with fixed external length scale \cite{smi93a}.
The goal of this paper is 
to theoretically understand the different flow regimes
which are to be expected, to explore the effect
of intermittency corrections on these regimes,
 and to ultimately initiate experiments.

Another motivation for the paper is to study the effect of a specified
type of forcing on turbulence. In most theoretical studies on turbulence a
Gaussian random noise, acting on the largest length scales, is assumed.
Only recently experimentalists started to systematically vary the type
of forcing \cite{lab96,cam97b}. 
This
work is a further step towards
the analysis of a more specific type of forcing. 

To define the model, we have to (i) calculate the energy input during the kick
and (ii) know how the energy is dissipated in the time
$\Delta t$ between successive kicks.

{\underline{(i) Kick:}}
As an illustration, consider plane shear flow in the 1-direction; the flow is
sheared in the 3-direction.
The width of the channel is $L$, the velocity of the upper plate is $U$, the
lower one is at rest. The average energy dissipation rate can be
calculated to be
(see e.g.\ \cite{geb95}) 
\be
\eps = -{U\over L} \left( \la u_3 u_1 \ra_A -\nu \partial_3 \la u_1 \ra_A
\right),
\label{eq1}
\ee
where $\la\ra_A$ denotes the average over the x-y-plane and $\nu $ is the
viscosity.
For laminar flow the first term in the bracket is zero and
the second one is $-\nu U/L$. In turbulent flow
in the middle of the channel 
the second term on the rhs will hardly
 contribute for large enough Reynolds numbers. 
The first term
 represents the total turbulent flow energy $E$, order of magnitude wise.
Therefore, in general,
\be
\dot E(t) = -\eps (t) \sim {U\over L} E(t) + {U^2 \over L^2} \nu .
\label{eq2}
\ee
Imagine now a short intense kick of time $\Delta t_{kick} \ll \Delta t$
on the flow by rapidly moving the upper plate with $U$.
After this kick the initial energy $E_0$ increases according to (\ref{eq2})
\be
E_1 = E_0 +\left( E_0 + { U\nu \over L}\right) {U\over L} \Delta t_{kick}, 
\label{new1}
\ee
where we have assumed $\Delta t_{kick} \ll L/U$. 
Assume isotropic turbulence in the flow center and define a Reynolds number
{\footnote{Note that this is not the standard definition and
gives lower values for the laminar-turbulent transition than what one
is used to.}}
\be
Re(t) = {L u_{1,rms}(t) \over \nu} = \sqrt{2\over 3}{L \sqrt{ E(t)}\over \nu}.
\label{eq3}
\ee
Then eq.\ (\ref{new1}) translates to
\be
Re_1 = Re_0 \sqrt{ 1+ 2A + {Re_{lam}^2 \over Re_0^2}}
\label{eq4}
\ee
with the dimensionless kicking strength 
$A = {1\over 2} \Delta t_{kick} U/L
 \ll 1 $ 
and the ``laminar'' Reynolds number $Re_{lam} = {2\over 3} \Delta t_{kick}
U^2/\nu $. We choose this name because for very small $Re_0$ we have
$Re_1 = Re_{lam}$. For very large $Re_0 \gg Re_{lam}$ we have
$Re_1 = (1+A) Re_0$.
Shear flow and the
``derivation'' of
eq.\ (\ref{eq4}) are only thought of as a motivation;
there will be many other experimental situations where the energy
input roughly corresponds to a law of type (\ref{eq4}).

{\underline{(ii) Decay:}}
In ref.\ \cite{loh94a} we calculated how the turbulent activity decays within a
time $ t$ for flow with fixed external length scale $L$. The
calculation was based on Effinger and Grossmann's variable range mean field
theory of turbulence \cite{eff87} in which viscous subrange and inertial
subrange can be treated equally well. The result of ref.\ \cite{loh94a} is that
for given initial Reynolds number $Re_i$ the time dependence of $Re( t)$
(defined as in eq.\ (\ref{eq3}))
is determined by the inverse function of
\be
{ t (Re) \over \tau } = {3\over \cinf} \left[ F(Re) - F(Re_i)\right]
\label{eq5}
\ee
where $\tau = L^2 /\nu $ and $F(Re)$ is given by
\begin{eqnarray}
F(Re) &=& {1\over 2Re^2} \left\{ -\gamma + \sqrt{\gamma^2 + Re^2} \right\}
\nonumber \\
&+& {1\over 2\gamma} \log
\left\{ {\gamma + \sqrt{\gamma^2 + Re^2} \over Re} \right\}.
\label{eq6}
\end{eqnarray}
$\cinf = (6/b)^{3/2}$ (in this theory) is the dimensionless energy
dissipation in the large Reynolds number limit,
$\gamma = 9/\cinf$, and $b$ is the Kolmogorov constant \cite{my75} which is the
only free parameter in the theory of ref.\ \cite{loh94a}.
From experiment
\cite{sre95}
$b=6.0$, a value which we take in all calculations here. Consequently,
$\cinf = 1.0$ and $\gamma = 9.0$. 
Rather than the Reynolds number eq.\
(\ref{eq3}) one could also give the Taylor-Reynolds number $\rel$
\cite{loh94a}, 
\be
\rel = \sqrt {15 Re^2 \over \cinf \left( \gamma + \sqrt{\gamma^2
+ Re^2}\right)}.
\label{eq7}
\ee

Eqs.\ (\ref{eq4}) - (\ref{eq6}) with $t=\Delta t$
(the time between successive kicks) 
fully define the present model. The two main
physical
parameters in the model are the kicking strength $A$ and the kicking frequency
$f=1/\Delta t$. The third physical parameter is $Re_{lam}$, the minimal
Reynolds number after a kick. We pick $Re_{lam } = 1$ throughout.

%caption1
\begin{figure}[htb]
\setlength{\unitlength}{1.0cm}
\begin{picture}(6,8.5)
\put(-0.0,1.0)
{\psfig{figure=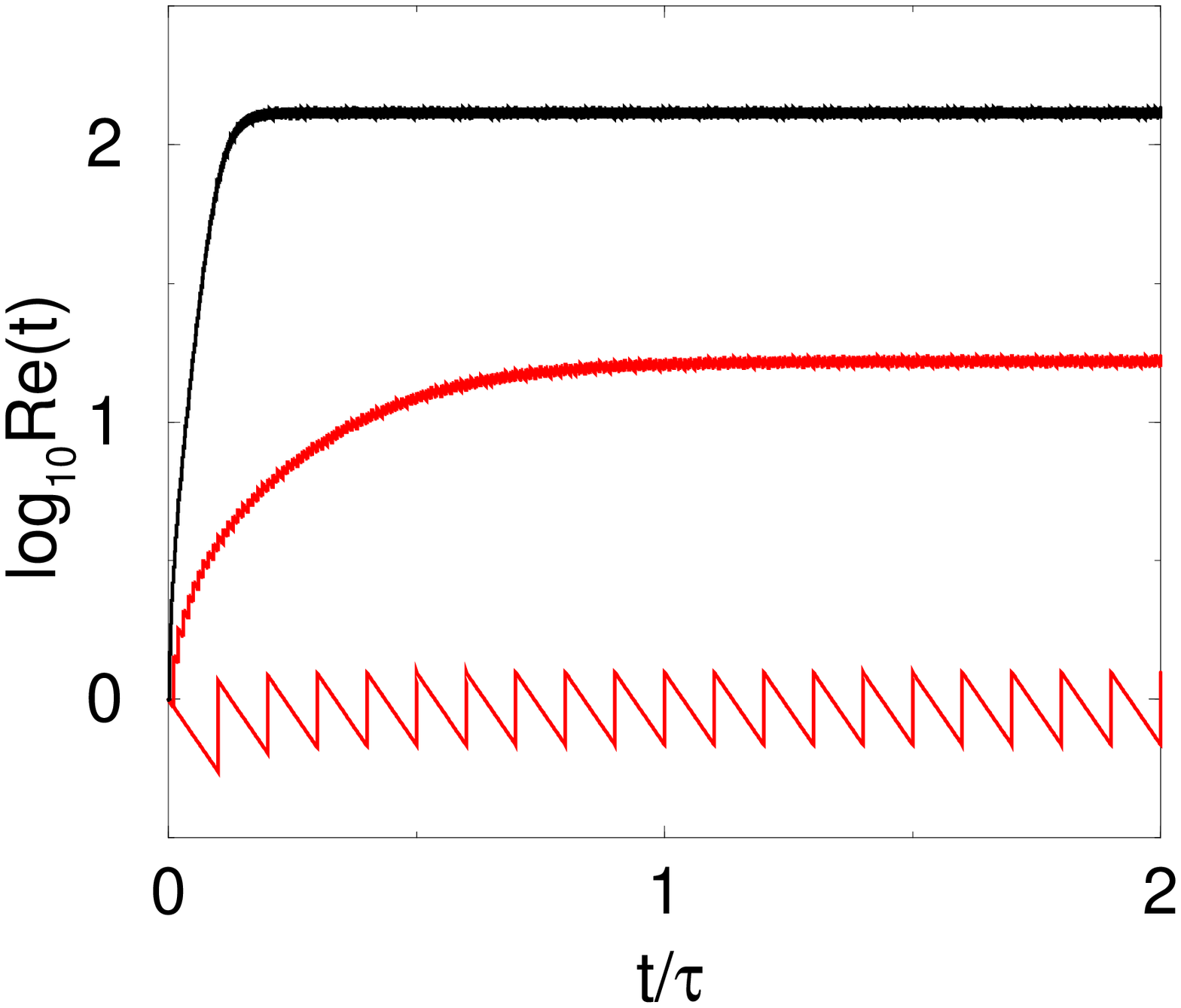,width=9cm,angle=-90}}
\end{picture}
\caption[]{
Time series $Re(t)$ for fixed $A=0.1$ and three kicking frequencies
$f=10/\tau$,
$f=100/\tau$, and $f=500/\tau$, bottom to top. 
The
corresponding (upper) saturation levels are
$Re_u^{sat} = 1.3 $,
$Re_u^{sat} = 17.3$, and
$Re_u^{sat} = 136$, respectively. 
}
\label{reoft}
\end{figure}

Figure \ref{reoft} shows $Re(t)$ for fixed $A$ and three different kicking
frequencies $f$ for the initial Reynolds number $Re_0 = Re_{lam} =1$. 
During each
cycle there is a kick $Re_0 \to  Re_1$ 
according to eq.\ (\ref{eq4}) 
and a subsequent decay according to
eq.\ (\ref{eq5}).
Overall, there is
growth up to some saturation level $Re^{sat}(A,f)$, achieved after
$t_{sat} (A, f)$. In this saturation state energy input and loss through decay
in $\Delta t$ balance and the degree of excitation fluctuates between
a lower saturation level
$Re_l^{sat}$ and an upper saturation level
\be
Re_u^{sat} = Re_l^{sat} \sqrt{1+2A +(Re_{lam}/Re_l^{sat})^2}.
\label{upper}
\ee

\begin{figure}[htb]
\setlength{\unitlength}{1.0cm}
\begin{picture}(6,8.0)
\put(-0.,1.0)
{\psfig{figure=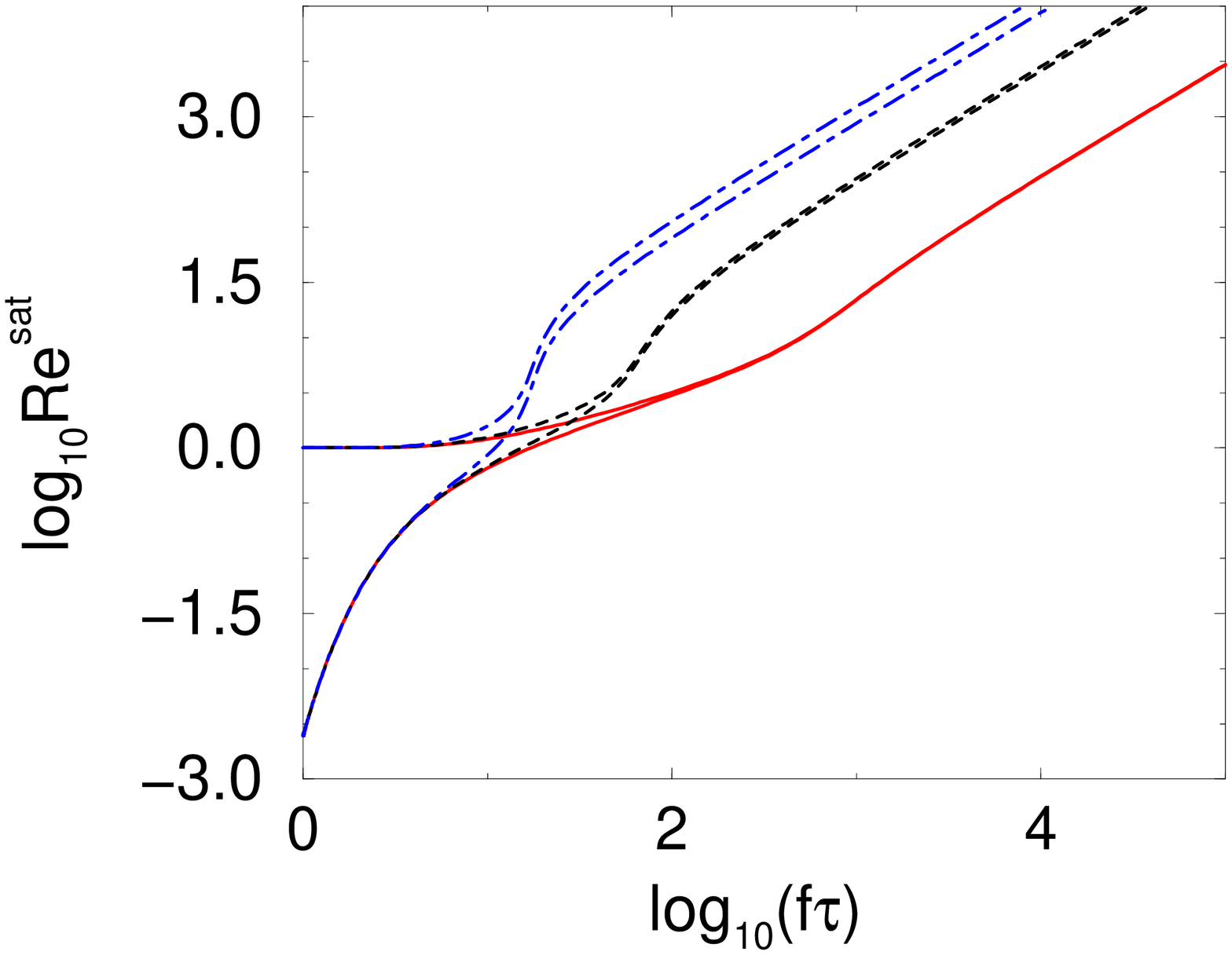,width=9cm,angle=-90}}
\end{picture}
\caption[]{
Saturation level $Re_l^{sat}$ (lower curve of pair)
and $Re_u^{sat}$ (upper curve of pair) 
as a function of $f$ for three
different kicking strengths
$A=0.01$ (solid),
$A=0.1$ (dashed), and 
$A=0.5$ (dashed-dotted), 
bottom to top. Below $Re_u^{sat} \sim \gamma =9.0$ the
excited state is laminar, above $Re_u^{sat} \sim 9$ it is turbulent. 
}
\label{resat}
\end{figure}

\begin{figure}[htb]
%\begin{figure}[p]
\setlength{\unitlength}{1.0cm}
\begin{picture}(6,7.5)
\put(0.0,1.0)
{\psfig{figure=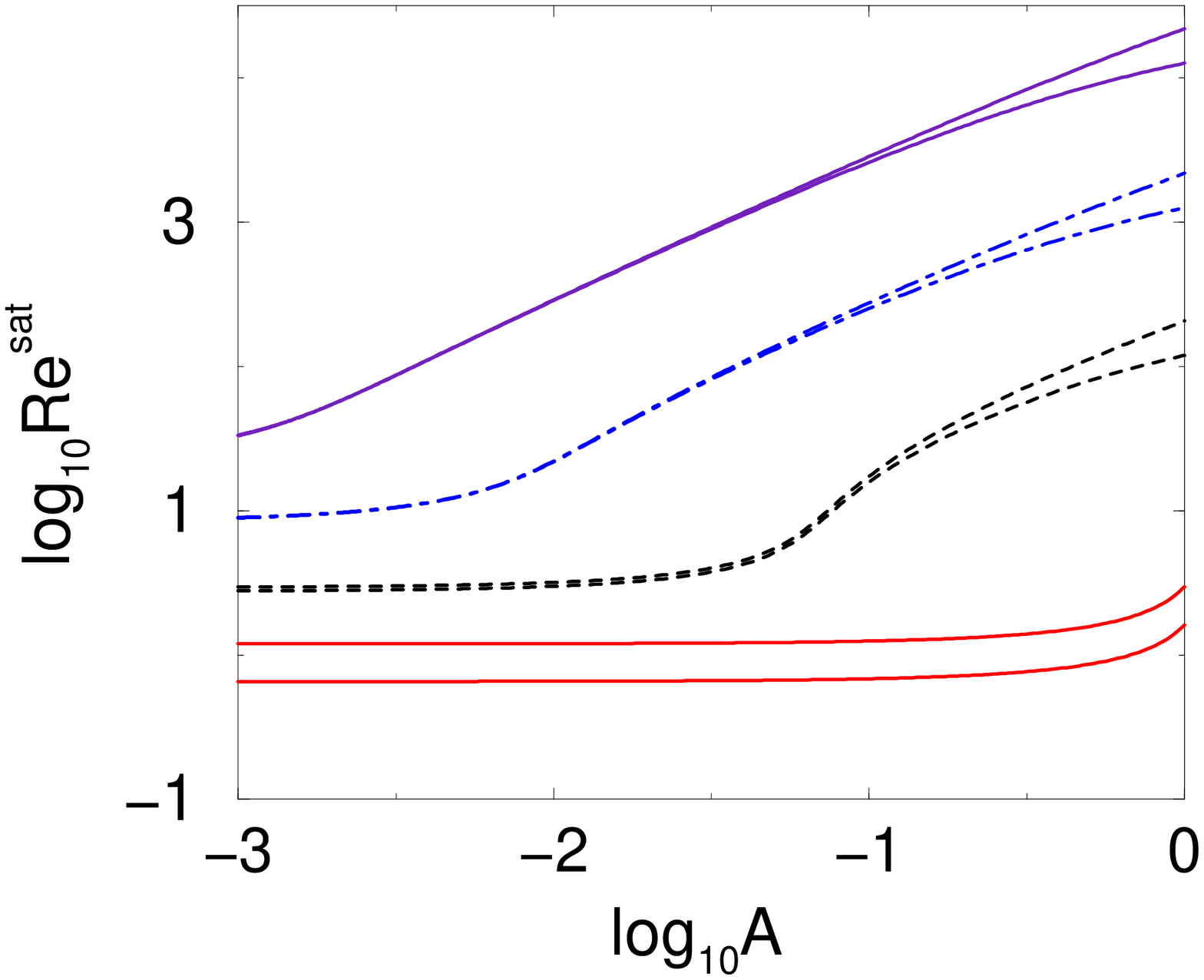,width=9cm,angle=-90}}
\end{picture}
\caption[]{
$Re_u^{sat}$ and 
$Re_l^{sat}$ 
for $f=10$, $10^2$, $10^3$, $10^4$, bottom to top. 
The increasing difference between
$Re_u^{sat}$ and 
$Re_l^{sat}$ at the right edge of the figure has its origin in
the breakdown of the requirement $A\ll 1$. 
}
\label{resat_of_a}
\end{figure}

The (lower) level of saturation
$Re_l^{sat}(A,f)$ is given through the implicit 
equation
\be
{1\over \tau f} = {3\over \cinf} \left[ 
F\left( {Re_l^{sat} }\right)
- F\left(
 Re_l^{sat} \sqrt{1+2A +\left({Re_{lam}\over Re_l^{sat}}\right)^2}
\right)\right].
\label{eq8}
\ee
For large
 $Re_u^{sat}$, $Re_l^{sat} \gg \gamma$ one has the explicit result
\be
Re_u^{sat} = {3\tau f \over \cinf } (\sqrt{1+2A} -1) 
\approx {3\tau \over \cinf } Af.
\label{eq9}
\ee
In figure \ref{resat} we show a log-log plot of $Re_{l,u}^{sat}$ as a function
of $f$ for different kicking strength $A$. Two  regimes are
seen:
(i) For small $Re_{l,u}^{sat} \aleq \gamma$
a laminar regime 
in which after the decay
the kick always brings back the level of excitation to the
laminar value $Re_{lam }=1$.
(ii) 
For large $Re_{l,u}^{sat}\ageq \gamma$
we have  a turbulent scaling regime with 
$Re_u^{sat} \propto f$. 
The transition from the laminar to the turbulence regime
takes place around
\be
f_{trans} = {3\over \tau A}
\label{transition}
\ee
 Similarily, 
these two regimes are also seen in 
fig. \ref{resat_of_a} where we plotted 
$Re_{l,u}^{sat}$ vs $A$. 
In the turbulent regime 
 for large $Re_{l,u}^{sat}$ it is 
$Re_{l,u}^{sat} \propto A$.

%caption1

%caption1

We now come to the important question of 
how 
intermittency effects \cite{fri95} change the exponents calculated within
this mean field theory. 
In refs.\ \cite{gro95}
intermittency effects have been included into the mean field theory
of ref.\ \cite{loh94a} on a phenomenological basis. 
One possibility for their effect is  that the dimensionless
energy dissipation rate $c_\eps$
becomes slightly Reynolds number dependent even
in the large Reynolds number limit \cite{gro95},
$c_\eps \propto Re^{-\kappa}$,
with $\kappa = (9/8) \delta \zeta_2 /(1+3\delta\zeta_2/8)$. Here,
$\delta\zeta_2 \approx 0.03$ is the experimentally found deviation from
the classical scaling
exponent $\zeta_2 = 2/3$ of the second order velocity structure function.
The consequences of this small ($\kappa \approx 0.03$) scaling correction
can straightforwardly be embodied in the present mean field approach
to periodically kicked turbulence. The  result is that in the turbulent
regime the saturation level now obeys
\be
Re_{l,u}^{sat} \propto (Af)^{1/(1-\kappa)}
\label{interm}
\ee
rather than eq.\ (\ref{eq9}).
Equation (\ref{interm}) may
offer a
new and independent way to experimentally determine
intermittency exponents. 

Another effect related to intermittency
is the following: 
We expect the total energy to build up 
again and again over time scales larger than $\Delta t$
and then to suddenly drop because a energy pulse is traveling downscale.
Such behavior has  been observed in numerical simulations 
of periodically remeshed homogeneous shear flow \cite{pum96}
and in simulations of periodically kicked shell modells of turbulence
[results to be publihsed]. 
As based on a mean field theory the model of this paper
is only applicable to the
mean energy and not to these fluctuations.

We now come back to an experimental realization.
It will be easier to 
perform experiments in a closed system 
rather than in a channel flow.
A particular suited experimental setup for periodically kicked
turbulence would be the flow in a cylinder between two counterrotating
disks \cite{lab96,zoc94}. Also, Rayleigh-Benard convection
may be well suited. Here, as an example we take Taylor-Couette flow 
\cite{lew99}: 
If the radii of the inner and outer cylinder are
similar, the energy input will still roughly follow eq.\ (\ref{eq4}).
Take water as fluid which has $\nu = 10^{-6}m^2/s$ and take $L=1cm$.
Then $\tau =100s$. Realistically achievable kick-strengths would be
$A= \Delta t_{kick} U /(2L) \sim  0.1s \cdot 0.1 m/s / (2\cdot 1cm) = 0.5$,
e.g., the whole range $A<1$ in which the theory is applicable. 
For this value the 
scaling regime sets in at 
$f_{trans} = 0.06Hz$. 
Roughly two  decades of scaling are necessary to explore
intermittency corrections
and to distinguish between eqs.\ (\ref{eq9}) and (\ref{interm}). 
I.e., one has to go up 
to frequencies of around $6Hz$ which should be achievable. 
At these relatively large kicking frequencies measurements
can only reveal {\it instantaneous} values. To get statements
on the averaged quantities dealt with in this paper, ensemble
averaging is necessary. This is best done by repeatedly probing
the flow at some fixed phase after the respective kick. Averaging
over the results at phase $0^+$ will give $Re_u^{sat}$, averaging 
over the results at phase $\Delta t^-$ will give $Re_l^{sat}$, etc.

We suggest to perform a periodically kicked turbulence 
experiment and to measure $Re^{sat}$ as a function of both $A$ and $f$.
To our knowledge it  would be one of the first
ones with some active control on the type of forcing.
Immediate questions to ask
are:
Does the level of saturation for large $Re$ indeed only depend on the
product $Af$ as suggested by eqs.\ (\ref{eq9}) and (\ref{interm})
or do boundary effects carry on into the strongly turbulent central regime and
cause a more subtle relation? If so, an application of the so commonly used
volume forcing for turbulence becomes more questionable. 
Do the (scaling) relations between the quantities
introduced in this paper, e.g., $Re_{l,u}^{sat} (f, A)$ 
offer a
new way to measure intermittency effects?
What modifications arise if forcing and decay do not decouple
as assumed in this simple model?

\vspace{0.5cm}

\noindent
{\bf Acknowledgements:}
The author thanks L. Biferale, M. Brenner, B. Eckhardt, 
and F. Toschi for
discussions. 
The work is part of the research  program of the Stichting voor 
Fundamenteel Onderzoek der Materie (FOM), which is financially supported 
by the Nederlandse  Organisatie voor Wetenschappelijk Onderzoek (NWO).
This research was also supported in part 
by the European Union under contract HPRN-CT-2000-00162 and 
by the National Science
Foundation under Grant No. PHY94-07194 and we thank the Institute
of Theoretical Physics in Santa Barabara for its hospitality.

\vspace{0.5cm}

\noindent
 e-mail address:
lohse@tn.utwente.nl

%\bibliography{literatur,sl_literatur}

\end{document}